\documentclass[useAMS,usenatbib]{mn2e}

\usepackage{verbatim}
\usepackage{graphicx}
\usepackage{amssymb}
\usepackage{amsmath}
\usepackage[figuresright]{rotating}
\newcommand{\mnras}{MNRAS}

\newcommand{\aap}{A\&A}
\newcommand{\prd}{PhRvD}
\newcommand{\apj}{ApJ}
\newcommand{\aj}{AJ}
\newcommand{\physrep}{PhR}

\title[Halo Mass Functions
 in  Early Dark  Energy   Cosmologies]{Halo Mass Functions
 in  Early Dark  Energy   Cosmologies}  \author[Francis,  Lewis  \&  Linder]
 {Matthew                J.                Francis$^{1}$\thanks{Email:
 mfrancis@physics.usyd.edu.au},  Geraint   F.   Lewis$^{1}$  and  Eric
 V. Linder$^{2}$  \\ $^{1}$ School  of Physics, University  of Sydney,
 NSW  2006,   Australia\\  $^{2}$   Berkeley  Lab  \&   University  of
 California, Berkeley, CA 94720, USA }

\begin{document}

\date{}

\pagerange{\pageref{firstpage}--\pageref{lastpage}} \pubyear{2007}

\maketitle

\label{firstpage}

\begin{abstract}
We examine  the linear density contrast at  collapse time, $\delta_c$,
for  large-scale  structure  in  dynamical  dark  energy  cosmologies,
including  models  with  early  dark  energy.   Contrary  to  previous
results, we find  that as long as dark energy  is homogeneous on small
scales,  $\delta_c$  is  insensitive  to dark  energy  properties  for
parameter  values fitting current  data, including  the case  of early
dark energy.   This is significant since using  the correct $\delta_c$
is crucial  for accurate Press-Schechter  prediction of the  halo mass
function.  Previous results  have  found an  apparent  failing of  the
extended  Press-Schechter  approach   (Sheth-Tormen)  for  early  dark
energy.  Our calculations demonstrate that with the correct $\delta_c$
the  accuracy   of  this  approach   is  restored.   We   discuss  the
significance of  this result  for the halo  mass function  and examine
what dark energy  physics would be needed to  cause significant change
in $\delta_c$, and the observational signatures this would leave.
\end{abstract}


\long\def\symbolfootnote[#1]#2{\begingroup%
  \def\thefootnote{\fnsymbol{footnote}}\footnotetext[#1]{#2}\endgroup} 

\symbolfootnote[1]{Research
  undertaken  as part  of  the Commonwealth  Cosmology Initiative  (CCI:
  www.thecci.org),  an  international  collaboration  supported  by  the
  Australian Research Council}

\section{Introduction}

Observations  of Supernovae Ia  \citep{riess98,perl99,kowalski08}, the
Cosmic  Microwave   Background  (CMB)\citep{wmap5}  and   large  scale
structure  \citep{2dfcosmo}  demonstrate  that  the expansion  of  the
Universe is  accelerating. Discovering the physics of  the dark energy
thought  to  be  driving this  phenomenon  is  a  key goal  of  modern
cosmology. Observations that probe  the non-linear growth of structure
are sensitive to the entire history  of the Universe and are a crucial
element in attempts to measure the evolution of dark energy properties
with time.

The abundance of collapsed structures  as a function of mass, the Halo
Mass  Function (HMF),  is an  important statistic  that  is measurable
through strong  lensing statistics \citep{lenseArcs},  galaxy redshift
surveys \citep{clusters} and  X-ray \citep{xray} detection of clusters
and  future cluster  surveys utilising  the  Sunyaev-Zel'dovich effect
signature  in  the CMB  \citep{plank}.   Accurate  estimation of  this
statistic as a function of cosmology is therefore required in order to
extract the maximum amount of information from observations.

Determining the  HMF of cosmological  models can be  a computationally
expensive  task   requiring  many  large   N-body  simulations,  since
non-linear  gravitational   structure  growth  cannot   be  calculated
analytically.  Therefore, simulation  calibrated tools for rapidly and
accurately generating  the observational signatures of  a wide variety
of models  are essential.  However, given  the theoretical uncertainty
surrounding the physics  of dark energy, tools for  predicting the HMF
must be valid for dark  energy models models as generally as possible,
to  avoid  detailed computation  for  every  one  of the  plethora  of
possibilities.

Current methods for  estimating the HMF fall into  two main categories
(see \citet{cooray}  for a  review).  The first  are methods  based on
\citet{PressSch} theory  that relate the density  of collapsed objects
of a given  mass to variance of the density  field on scales enclosing
that  mass in  the  mean, $\sigma^2(M)$,  and  a threshold  parameter,
$\delta_c$, determining  the linear overdensity  required for collapse
by a given redshift.  The leading approach based on these ideas is the
\citet{ShethTormen} (hereafter  ST) mass function,  which incorporates
ellipsoidal  collapse (rather  than the  purely spherical  collapse of
Press-Schechter theory)  and has  free parameters that  are simulation
calibrated.

The second type  of mass function fitting approach  is to fit directly
the multiplicity function
\begin{equation}
f(\sigma) = \frac{M}{\bar{\rho}}\frac{dn(M,z)}{d\textrm{ln}\sigma^{-1}}
\end{equation}
by a universal function of  the variance
\begin{equation}
\sigma^2(M) = \frac{1}{2\pi^2}\int_0^\infty k^2 P(k) W^2(k,M) dk
\end{equation}
in  which  $P(k)$  is  the  power spectrum  of  density  fluctuations,
$\bar{\rho}$ is the  mean matter density, and $W(k,M)$  is the Fourier
transform of a spherical top  hat function with a radius that encloses
the mass  $M$ at the  mean density of the  universe. \citet{Jenkins01}
(hereafter  J01)   found  a   universal  function  of   $\sigma$  from
simulations of $\Lambda$CDM, matter  dominated and open universes that
fits  the multiplicity  function for  all of  these  cosmologies.  More
recently  \citet{Warren} have  found  a similar  formula that  fitted
their simulation results slightly better.

Since there are many  different dark energy models currently proposed,
it has not proved practical  to include dark energy variation into HMF
fitting work, and currently  available formulas are calibrated only to
$\Lambda$CDM and  matter only  cosmologies.  However, the  J01 formula
has been demonstrated via N-body  simulations to be valid (at least at
the   $20\%$    level)   for    an   evolving   dark    energy   model
\citep{LinderJenkins} and the ST formula  was also found to agree with
evolving  dark energy  models in  \citet{klypin}, indicating  that the
universality of  gravitational collapse appears to extend  to at least
some dark energy models.

One particularly  interesting class of evolving dark  energy models is
the concept  of early  dark energy (hereafter  EDE).  In  these models
\citep{Wetterich04,DoranRobbers06},  dark energy has  a non-negligible
energy fraction  through the entire  course of cosmic  history, rather
than being important only at  late times as in the $\Lambda$CDM model.
In  an investigation  of the  HMF in  EDE  cosmology, \citet{bartDW06}
(hereafter BDW)  calculated, using the ST approach,  that the presence
of  EDE  leads  to  a  significant enhancement  of  the  abundance  of
collapsed  objects  relative  to  $\Lambda$CDM, particularly  at  high
redshift.  However,  in a recent study \citep{francis},  we found that
the J01 and Warren et al.  mass functions, in contrast, predicted much
less difference  between EDE  and $\Lambda$CDM and  N-body simulations
agreed with this result.  The ST results using the BDW $\delta_c$ were
clearly ruled out by simulation data.

Why  did the  Press-Schecter (Sheth-Tormen)  approach appear  to fail?
While  the multiplicity  function is  expressed directly  in  terms of
$\sigma$,  the  Press-Schechter  approach involves  $\delta_c/\sigma$,
suggesting that $\delta_c$ could be the source of the discrepancy. BDW
calculated that  $\delta_c$ is significantly altered  in EDE cosmology
compared to $\Lambda$CDM and this difference in $\delta_c$ resulted in
the difference in mass  functions. \citet{gross} noted that if instead
of   using  spherical   collapse  arguments,   a  constant   value  of
$\delta_c=1.689$ is  assumed for all cosmologies at  all redshifts the
basic agreement between the ST and J01 mass functions is restored.  In
this  work we investigate  this issue  further, re-examining  the root
calculation of $\delta_c$.

In  Section  \ref{calcdc},   we  re-examine  methods  for  calculating
$\delta_c$  and advocate  a more  accurate technique  for  dark energy
models, especially  for early dark  energy. Given our  calculations of
$\delta_c$,  in   Section  \ref{halostuff}  we   determine  halo  mass
functions from  the ST approach and  compare to the  J01 approach.  We
also  discuss  the implications  for  mass  functions  in dark  energy
cosmologies more generally.  Finally  in Section \ref{disc} we discuss
the dependence  of $\delta_c$  on dark energy  and suggest  under what
conditions we  might expect a significantly different  value from that
of $\Lambda$CDM and the observational consequences.

\section{Computing the Linear Density Contrast}\label{calcdc}

The  picture  of  non-linear  growth  of  structure  involves  density
perturbations  growing  in amplitude,  initially  by  a linear  growth
factor, then  achieving sufficient  density contrast to  separate from
the  Hubble expansion  and to  collapse, increasing  its density  in a
non-linear manner.  One  can calculate the level to  which the density
contrast  would have grown  in linear  theory by  collapse time,  as a
convenient parameter  (and an essential  ingredient in Press-Schechter
formalism),  though  the true,  non-linear  density  contrast is  much
larger.

The linear density contrast at collapse, $\delta_c$, is defined by
\begin{equation}\label{dcdef}
\delta_c = \lim_{a\to0}\left[\Delta(a)-1\right]\frac{D_+(a_c)}{D_+(a)}
\end{equation}
where  $\Delta$   is  the  overdensity,   $\rho/\bar{\rho}$,  of  some
spherical region of the universe that collapses at scale factor $a_c$,
$D_+(a)$  is the  linear  growth factor.   This parameter,  therefore,
quantifies the linear growth from the early universe until $a_c$ of an
overdensity known  to collapse under  non-linear growth by  that time.
For   matter   dominated   cosmologies,   $\delta_c=1.686$   and   for
$\Lambda$CDM it  becomes a weak  function of cosmology,  retaining the
matter dominated value  at high redshift and dipping  only slightly by
redshift zero.

In order to  calculate $\delta_c$, we need to  solve the linear growth
equations  to obtain  $D_+(a_c)/D_+(a)$.   We also  need  to find  the
overdensity  in the  early  universe  that leads  to  collapse of  the
perturbation  at  the exact  desired  scale  factor  $a_c$. The  usual
approach  (see e.g.  \citet{peebles}) is  to solve  simultaneously the
Friedmann  equations  for the  background  universe and  perturbation,
treating  the   perturbation  as  a  closed  universe.    It  is  also
traditional to normalise these  equations to turnaround, when the time
derivative of the perturbation radius is zero.  Following the notation
of  BDW, we  therefore  need  to solve  the  following equations  (for
simplicity we take a flat background universe)

\begin{equation}\label{fred1}
\dot{x}= \sqrt{ \frac{\omega}{x} + \lambda x^2 g(x)}
\end{equation}

\begin{equation}\label{fred2}
\ddot{y}= -\frac{\omega \zeta}{2 y^2} - \frac{1+3w(x)}{2}\lambda g(x)y
\end{equation}
where $x\equiv a/a_{ta}$ and $y\equiv R/R_{ta}$, and $R$ is the radius
of  the  perturbation  (which   is  assumed  to  be  spherical).   The
dimensionless  density  parameters  of   matter  and  dark  energy  at
turnaround  are  $\omega$   and  $\lambda$  respectively  and  $\zeta$
quantifies the  overdensity at turnaround;  $w(x)$ is the  dark energy
equation of state and $g(x)$  is the dark energy density normalised to
the turnaround  value.  The dots indicate derivatives  with respect to
the  time parameter  $\tau \equiv  H_{ta}t$  where $H$  is the  Hubble
parameter and $t$  is cosmic time.  Finding the  value of $\zeta$ that
ensures  collapse of  the perturbation  at the  required  scale factor
$a_c$ requires a numerical search.

The overdensity, $\Delta(x)$, can be defined via
\begin{equation}
\Delta(x)=\frac{\zeta x^3}{y^3}
\end{equation}
and  therefore,   once  $\zeta$  is  found,  the   behaviour  of  Eqs.
(\ref{fred1}) and (\ref{fred2}) determines the size of the overdensity
at early times.

In BDW, an approximate solution for the overdensity at early times was
derived of the form
\begin{equation}
\Delta(x)=1+\frac{3}{5}F\zeta^{1/3}x 
\end{equation}
where there are two solutions for  $F$, one for the case when the dark
energy equation of state at early times, $w_i \equiv \lim_{x\to0} w(x)
<-1/3$, and  one for $w_i >  -1/3$. $\Lambda$CDM is an  example of the
first case, and the solution results in
\begin{equation}\label{FL}
F_{\Lambda CDM} = 1 + \frac{\lambda}{\omega \zeta}.
\end{equation}
The second case includes the \citet{Wetterich04} EDE model examined in
BDW, with 
\begin{equation}\label{FEDE}
F_{EDE} = 1 - \frac{\Omega_e}{1-\Omega_e}
\end{equation}
where $\Omega_e$  is a parameter of the  \citet{Wetterich04} EDE model
quantifying  the fractional dark  energy density  at early  times.  In
this  model,   as  $\Omega_e  \to  0$,  the   cosmology  converges  to
$\Lambda$CDM, and hence we should expect that these two solutions also
converge. However this does not  occur, in this limit Eq. (\ref{FEDE})
converges to unity and not to Eq. (\ref{FL}).

Moreover,  the non-negligible  presence of  dark energy  in  the early
universe should,  relative to $\Lambda$CDM,  slow the collapse  of the
perturbation due  to the  lower matter clustering  source term  in the
early universe and later the  higher expansion rate.  We should expect
then, that for  the same collapse scale factor  $a_c$, the addition of
EDE should require increasing the overdensity in the early universe in
order to  compensate for the  slower non-linear growth.   However, the
solution for  $\Delta(x)$ in  the early universe  from BDW  predicts a
{\it decrease}  in the overdensity  for EDE relative  to $\Lambda$CDM.
These problems raise doubts about the accuracy of this solution.

We note that  a precise determination of the  initial $\Delta(a_i)$ is
needed in  order to accurately determine $\delta_c$.   We find typical
values of $\Delta(a_i)$ to be $\Delta(a_i) \simeq 1+3(a_i/a_c)$ where,
to  ensure numerical  convergence (results  independent of  the chosen
$a_i$) we  take $a_i <  a_c \times 10^{-4}$.   This means that  even a
small  error, say  of  order $10^{-3}$  in  $\Delta(a_i)$, is  greatly
amplified   when  going  to   $\Delta(a_i)-1$,  needed   to  calculate
$\delta_c$ in Eq. (\ref{dcdef}).  In this example the error induced in
$\delta_c$ would then be of order unity.

We   employ  a  different,   straightforward  approach   to  computing
$\delta_c$, purely numerically. We  still solve Eqs. (\ref{fred1}) and
(\ref{fred2}), however  rather than  searching for the  overdensity at
turnaround  numerically   then  scaling  this  back   to  early  times
approximately,   we  perform   the  numerical   search   starting  the
integration  at  early  times,  and  hence  search  directly  for  the
overdensity  $\Delta(a_i)$ that  causes collapse  of  the perturbation
($y\to 0$) by  the desired collapse scale factor  $a_c$.  Our approach
has several advantages:

\begin{itemize}
\item In  the previous  approach, a numerical  search must be  made to
find the overdensity  at turnaround before this can  be scaled back to
the  early universe  using  an approximate  solution. Performing  this
search at the  early time instead avoids the  need for the approximate
scaling back without adding to computation time.

\item Finding  the overdensity  at turnaround implicitly  assumes that
the  rise  and  fall times  of  the  radius  of the  perturbation  are
equal. This  is not true in general  dark energy cosmologies\footnote{
From  Eq.(\ref{fred2}) we see  that the  symmetry in  the perturbation
rise and fall times holds when the coefficient of $y$ in the last term
is time  independent (then it gives the  ``development angle" solution
$y\sim\cos^{2/3}[C(t-t_a)]$,  where  $C$  is  constant,  cf.\  Peebles
1980).  The coefficient is indeed  time independent in three cases: 1)
when $\lambda=0$  (pure matter universe),  2) when $g(x)$  is constant
(which  implies  $w(x)=-1$:  $\Lambda$CDM),  and 3)  when  $w(x)=-1/3$
(equivalent  to  spatial  curvature,  i.e.\ a  matter  plus  curvature
universe such  as OCDM).  These  were the cases most  researchers were
interested in  until the late 1990s,  and so symmetry  has passed into
the lore of linear collapse.  However, it is {\it not\/} true for dark
energy  cosmologies  with  $w  \ne  -1$!}. By  avoiding  reference  to
turnover, we avoid this approximation.

\item By solving the exact equations, we can quantify the magnitude of
any errors  introduced by  approximations that may  lead to  a simpler
functional  form for  $\Delta(x)$.   Without this  solution we  cannot
properly test the validity of approximations.
\end{itemize}

\begin{figure}
\includegraphics[
  scale=0.35,
  angle=-90]{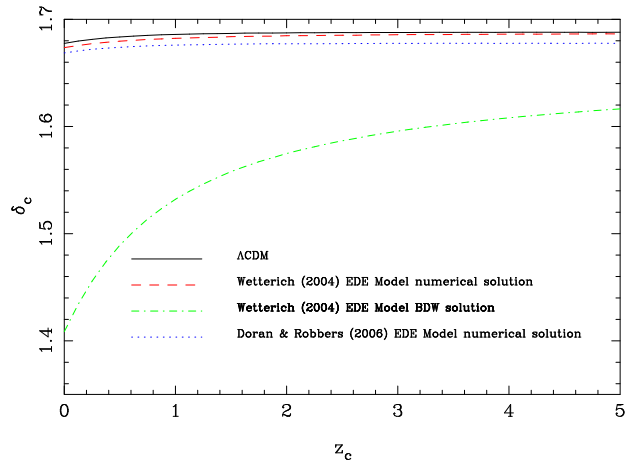}
  \caption{Linear density  contrast at collapse  time, $\delta_c$, for
  $\Lambda$CDM  and EDE  cosmologies.  The  $\Lambda$CDM  model (solid
  black  line), is  tracked closely  by our  numerical result  for EDE
  (dashed red line).  By contrast, the method of BDW (dot-dashed green
  line) predicts a significant  decrease in EDE cosmology.  Our result
  for the  \citet{DoranRobbers06} EDE model  (dotted blue line)  has a
  somewhat greater departure from  the $\Lambda$CDM value than for the
  \citet{Wetterich04}  model,   even  at  high   redshifts.   For  the
  \citet{DoranRobbers06} model, the  BDW approach gives $\delta_c\to0$
  (see   Francis   et  al.    2008),   so   we   do  not   plot   that
  case.}\label{dcplot}
\end{figure}

Results from our numerical calculation, compared to the method of BDW,
are shown in  Fig.~\ref{dcplot}.  The EDE model is  model (I) from BDW
with      parameters      $\Omega_m=0.325$,      $w_0=-0.93$,      and
$\Omega_e=2\times10^{-4}$.   This is compared  to a  flat $\Lambda$CDM
model with $\Omega_m=0.3$ as in BDW.  We also compare to different EDE
model,  proposed  by \citet{DoranRobbers06},  which,  compared to  the
\citet{Wetterich04} model,  has a greater  EDE density at  early times
relative to its $z\approx2$ value. We show our results for this model,
with $\Omega_m=0.3$,  $w_0=-1$ and $\Omega_e=0.05$  (note that current
data permits higher values of $\Omega_e$ in the \citet{DoranRobbers06}
model  than   the  \citet{Wetterich04}  model,   see  \citet{DRW}  for
details).
The key  result from Fig.~\ref{dcplot}  is that we  find $\delta_c$ in
EDE cosmology to be little  changed compared to $\Lambda$CDM.  This is
significant, and  in contrast  to previous results.   The implications
for  the ST mass  function are  discussed in  Section \ref{halostuff}.
The  reasons for breakdown  of the  BDW solution  are detailed  in the
Appendix.

The dependence of $\delta_c$ on EDE is fit by
\begin{equation}\label{fit}
\delta_c^{EDE}(a_c)= A + \left[b(1+w_0)+c\Omega_m-d\right]a_c-e\Omega_e
\end{equation}
where for  the \citet{DoranRobbers06} model,  $A=1.6905$, $b=-0.0183$,
$c=0.0264$, $d=0.0208$ and  $e=0.202$, and for the \citet{Wetterich04}
model,    $A=1.6899$,   $b=-0.0170$,   $c=0.0455$,    $d=0.0307$   and
$e=0.753$. Both  of these  fits are good  to $\sim0.1\%$ in  the range
$0.2<\Omega_m<0.4$,  $-1.2<w_0<-0.8$,  $0.1<a_c<1.0$  and  $0<\Omega_e
<0.05$   for   the   \citet{DoranRobbers06}  model   and   $0<\Omega_e
<1\times10^{-3}$  for   the  \citet{Wetterich04}  model.    By  taking
$\Omega_e=0$  and  $w=-1$,  Eq.   \ref{fit} returns  the  $\Lambda$CDM
result.

\section{Halo Mass Functions}\label{halostuff}

\begin{figure}
\includegraphics[
  scale=0.35,
  angle=-90]{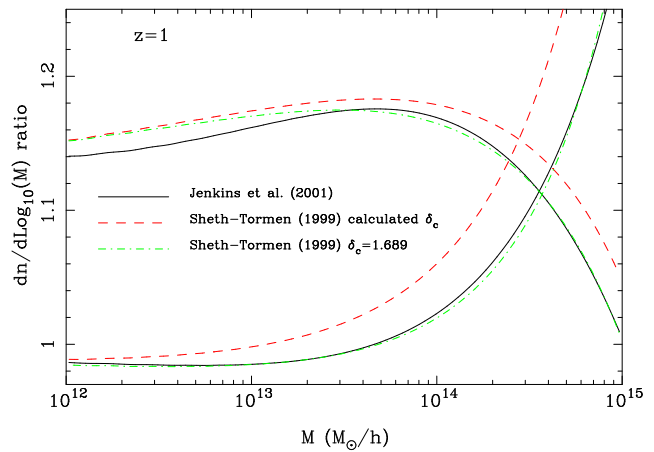}
  \caption{Comparison between the ST and J01 mass functions. The upper
  set of  curves are for  the \citet{Wetterich04} model and  the lower
  set the \citet{DoranRobbers06} model. The solid black line shows the
  calculated  ratio between  the EDE  and $\Lambda$CDM  using  the J01
  formula.  The red dashed line shows the same ratio using the ST mass
  function for both models  using our calculated values of $\delta_c$.
  The green  dot-dashed line shows the ST  ratio when $\delta_c=1.689$
  is used instead of the calculated values. The \citet{DoranRobbers06}
  mass functions  use linear  power spectra, $P(k)$,  for the  EDE and
  $\Lambda$CDM models  normalised to a common  $\sigma_8$ today, while
  the \citet{Wetterich04} mass functions use the models from BDW which
  are picked  from a Monte-Carlo  chain fitting current data  and have
  some   differences   in   normalisation  and   primordial   spectral
  index. }\label{haloplot}
\end{figure}

The  results on  the linear  collapse  parameter solve  a puzzle  from
\citet{francis}, where a marked difference between the ST and J01 mass
functions in EDE cosmologies  was highlighted, and \citet{gross} where
it was noted that if spherical  collapse is ignored and a common value
of  $\delta_c=1.689$ is  assumed instead  the agreement  between these
methods  is  restored.  For  the  ST  mass  function, using  spherical
collapse  arguments  for  $\delta_c$,  both studies  relied  upon  the
calculation  from BDW.   Using  instead the  method  outlined in  this
study, we have re-examined the ST  and J01 mass functions. As found in
\citet{francis},  the  EDE  mass  functions are  not  greatly  altered
compared to  $\Lambda$CDM at  $z=0$, however the  difference increases
with redshift.  The  EDE mass functions as a  ratio to $\Lambda$CDM at
$z=1$  for   the  same  models  as  Fig.~\ref{dcplot}   are  shown  in
Fig.~\ref{haloplot}.
This result demonstrates  that the basic agreement between  the ST and
the J01 mass functions is  preserved, even when the spherical collapse
motivated, cosmology dependent, $\delta_c$  is used. Thus not only the
form, but also  the conceptual basis of the ST  mass function is valid
for  EDE.   We note  that  at the  high  mass  end ($M\gtrsim  10^{13}
M_{\sun}/h$), the choice of calculating $\delta_c$ or holding it fixed
makes as big a difference as choosing between the ST and J01 formulas.
Neither  the simulations from  \citet{francis} nor  \citet{gross} have
sufficient accuracy at this high  mass end to make any clear judgement
about which choice better fits simulation data.

The success  of the J01  style mass functions,  that are blind  to the
growth history of the universe (as opposed to the instantaneous growth
factor), indicates  that indeed the abundance of  halos is insensitive
to  this.   When the  growth  history  {\it  is} considered,  via  the
alteration  of $\delta_c$  in the  ST mass  functions, small,  but not
insignificant, differences between the  ST and J01 predictions for the
relative mass  function in EDE  and $\Lambda$CDM emerge.   Future work
with simulations containing sufficient  volume to accurately probe the
high mass range could discriminate  between the two approaches to mass
function fitting and determine whether the growth history affects halo
abundances.

While  the abundance  of halos  is unaffected  by the  growth history,
\citet{francis}   found  that   non-linear  power   at   small  scales
($k\gtrsim1$)   is   increased   in   EDE  cosmologies   relative   to
$\Lambda$CDM. Since  this part of  the power spectrum is  dominated by
the  one-halo term  in  the halo  model  \citep{cooray}, the  internal
density  profile   of  halos  in   EDE  will  be  different   than  in
$\Lambda$CDM. This is also seen in the results of \citet{gross}.

\section{Discussion and Conclusion}\label{disc}

In this  letter we have demonstrated  that the ST  mass function, when
the correct $\delta_c$ is used, agrees with the J01 and \citet{Warren}
mass  functions. For  reasonable parameter  values, $\delta_c$  in EDE
models is not significantly altered compared to $\Lambda$CDM.  This is
in contrast to  previous results for EDE, but  agrees with analyses of
other     dynamical    dark     energy     models,    for     instance
\citet{2003ApJ...599...24M}.  From Eq.  (\ref{dcdef}), we can see that
$\delta_c$ is  defined by comparing linear to  non-linear growth.  If,
in  some cosmology,  we find  $\delta_c$ to  be  significantly altered
compared to some other model,  then this indicates that the difference
between  models must  be  altering the  linear  and non-linear  growth
differently.

What kind of cosmology would have a significantly different $\delta_c$
compared to $\Lambda$CDM?  Some new  physics must alter the linear and
non-linear growth  rates in  different ways compared  to $\Lambda$CDM.
One key assumption we have made in our analysis is that dark energy is
smoothly distributed  on the relevant length scales.   This means that
the  dark energy  density in  the background  universe and  within the
perturbation   are  not   evolved  independently.    If   dark  energy
perturbations were in fact important  on small scales, or for instance
if dark energy was non-minimally  coupled to dark matter then the dark
energy density within the  perturbation will evolve differently to the
background universe.  In this case  we would expect a more significant
alteration of $\delta_c$, although careful determination of the linear
growth  as  well  as  the  non-linear spherical  collapse  taking  any
coupling or  dark energy perturbations properly into  account would be
needed, see  for  instance  \citep{ManMota,dc1,dc2,dc3,dc4}.

As  found in  BDW, if  $\delta_c$  is postulated  to be  significantly
different  to the $\Lambda$CDM  value, then  large differences  in the
abundance  of collapsed  objects would  be  seen, over  and above  any
difference we  might anticipate based upon measurements  of the linear
growth  rate from the  CMB, weak  lensing and  the large  scale galaxy
power spectrum normalisation combined  with knowledge of the expansion
history from supernovae Ia data. Since some discrepancies in structure
measurements may exist \citep{fedeli}, it therefore remains a question
for further studies as  to whether such observations potentially point
to the non-linear  growth in our universe not  following the universal
form and hint at added physics for dark energy or dark matter.

\section*{Acknowledgments}

We thank  Carlos Frenk  and Ravi Sheth  for useful  conversations. GFL
acknowledges support from ARC  Discovery Project DP0665574.  This work
has been supported in part  by the Director, Office of Science, Office
of  High Energy  Physics,  of  the U.S.\  Department  of Energy  under
Contract No.\ DE-AC02-05CH11231.

\section*{Appendix: The BDW solution for $\Delta(\lowercase{x})$}\label{appendix}

The  calculation of  $\delta_c$ in  BDW makes  two key  assumptions in
order to  approximate the early  time evolution of the  overdensity in
EDE cosmologies.  In Eq.  (\ref{fred1}), the second  term is neglected
on the  assumption that  it is  small compared to  the first,  and the
second  order  differential  Eq.   (\ref{fred2}) is  converted  to  an
approximate first order equation.   These approximate equations, for a
flat universe, are
\begin{equation}\label{app1}
\dot{x} \simeq \left[\omega/x\right]^{1/2}
\end{equation}
\begin{equation}\label{app2}
\dot{y}\simeq\left[\frac{\omega \zeta}{y} - \omega\zeta + \frac{\zeta \Omega_e \omega}{(1-\Omega_e)y}\right]^{1/2}
\end{equation}
To examine  the effects that these approximations  introduce, we first
determine $\zeta$ using our  numerical solution for $\Delta(a_i)$, and
then numerically integrate Eqs.   (\ref{app1}) and (\ref{app2}) in the
early  universe.  If the  approximations are  good the  results should
match our exact calculation in the early universe.

There is an important point that must be made using our approach.  The
boundary  conditions of  $\dot{y}_i/y_i=\dot{x}_i/x_i$  at $a_i$  (the
perturbation starts off co-moving with the Hubble flow) ensure that we
have both a growing and decaying mode initially, such that $\delta_+ =
(3/5)\delta_i$ where  $\delta \equiv \Delta -1$. At  the initial time
$x_i$, we  have both growing and  decaying modes and hence  we need to
apply  the factor  of $3/5$  in order  to extricate  the  growing mode
only. The essence of the BDW approach is to approximate the early time
trends of  Eqs.  (\ref{fred1}) and (\ref{fred2}),  which should return
simply  the growing  mode.  As  expected from  the results  of Section
\ref{calcdc},   the   integration   using   Eqs.    (\ref{app1})   and
(\ref{app2}) does not match our  result, however it does reproduce the
solution  of BDW after  the decaying  mode has  dissipated, indicating
that these approximations cause  the difference compared to our result
(both  approximations contribute  comparable errors).   The subsequent
manipulations of these  equations in BDW are clever  but the damage is
already done.

\bsp

\label{lastpage}

\end{document}